PAPER

# Experimental and numerical study on *Following Streamer* mechanism for $SF_6$ streamer propagation induced by floating linear metal particles under negative applied voltage

**Zihao Feng[1,2,*], Liyang Zhang[3], Xiaobing Zou[2] and Haiyun Luo[2]**

[1] High Voltage Department, China Electric Power Research Institute, Beijing 100192, People's Republic of China
[2] Department of Electrical Engineering, Tsinghua University, Beijing 100084, People's Republic of China
[3] College of Mechanical and Electrical Engineering, Beijing University of Chemical Technology, Beijing 100029, People's Republic of China
[*] Author to whom any correspondence should be addressed.

**E-mail:** fengzh20@tsinghua.org.cn

## Abstract

A recently proposed *Following Streamer* mechanism (Feng *et al* 2025 *Phys. Rev. Appl.* **23** 064039) seeks to explain how floating metal particles induce $SF_6$ streamer propagation in the combined gap. This mechanism is derived from a two-dimensional (2D) axisymmetric fluid model, which has limitations in describing multiple streamer events in real-world three-dimensional (3D) scenarios. To further study the *Following Streamer* mechanism, we experimentally investigate the discharge morphology of $SF_6$ streamers induced by a floating linear metal particle under negative applied voltage. The results are then compared with those from 2D axisymmetric fluid simulations. The comparison reveals both consistencies and discrepancies. Regarding consistencies, experimentally observed features-such as streamer inception at both ends of the metal particle and the formation of subsequent luminous phenomena within the $SF_6$ streamer channel-are qualitatively consistent with the general idea of the *Following Streamer* mechanism. Regarding discrepancies, the experiments show a larger number of following streamers and off-axis propagation paths, which cannot be described in the 2D simulation. For scientific rigor, an extended conceptual model was proposed to improve the description of the *Following Streamer* mechanism, using 3D schematic for visualization.

## 1. Introduction

Floating linear metal particles are considered one of the most critical threats to $SF_6$ insulation in gas-insulated electrical equipment [1–4]. Understanding how these particles induce $SF_6$ combined-gap bridging-particularly during the streamer stage, which marks the onset of the breakdown process-is crucial for advancing fundamental knowledge and potential application. However, a precise understanding of the microscopic transient behavior involved still remains limited. Recently, a so-called *Following Streamer* mechanism [5], derived from plasma fluid simulations, has been proposed to explain how metal particles trigger $SF_6$ streamer propagation in the combined gap. The core arguments and limitations of the *Following Streamer* mechanism are summarized as follows:

(1) ***Arguments***: The *Following Streamer* mechanism argues that interaction between $SF_6$ plasma and metal particles-along with the particles" electrostatic induction-leads to a self-consistent redistribution of surface charges [5, 6]. This redistribution significantly enhances the local electric field at the particle tips, producing an effect similar to the rising edge of a pulsed voltage. As a result, linear metal particles can initiate streamer discharges simultaneously at both ends. After the formation of the primary streamer, new streamers can be triggered, named following streamers. These following streamers fundamentally represent multiple streamer events, which are argued to facilitate the $SF_6$ combined-gap bridging.

(2) ***Limitations***: The approach of [5] is based on a two-dimensional (2D) axisymmetric deterministic fluid model, which presents significant limitations: it cannot accurately represent the three-dimensional (3D) nature of the multiple streamer events. Specifically, in [5], all following streamers are constrained to initiate and propagate along the axis of symmetry. However, in practice, multiple streamer propagation often breaks this symmetry [7–11]. The resulting asymmetry in the electric field can destabilize streamer paths, leading to fully 3D discharge morphologies that the 2D model cannot capture. To rigorously describe the *Following Streamer* mechanism, experimental investigations capable of resolving realistic streamer morphologies are therefore important.

In this paper, we experimentally investigate $SF_6$ streamer morphologies induced by a floating linear metal particle under negative microsecond pulsed voltage, and compare the results with 2D axisymmetric plasma fluid simulations. First, we develop a floating control apparatus for sub-millimeter metal particles. A full description of this apparatus, along with the experimental setup and procedures, is provided in section 2. The error analysis and rationality assessment of the experimental system are detailed in section 3. Besides, a brief description of the numerical model is presented in appendix. Using an exposure time of 0.5 ns, we capture photographs of $SF_6$ streamers in the context of combined-gap bridging. The experimental observations are then compared with simulation results, revealing both consistencies and discrepancies. The consistencies include the existence of double-ended streamer inception and subsequent luminous phenomena within the $SF_6$ streamer channel. The discrepancies involve both the number of following streamers and the symmetry of their formation positions and propagation paths. A detailed analysis of these findings, along with a discussion of the physical mechanisms involved, is presented in section 6. Finally, based on these results, we propose an improved conceptual model describing $SF_6$ streamer propagation induced by floating linear metal particles under negative applied voltage in section 5.

## 2. Experimental setup and method

### 2.1. Floating control apparatus

The key element of this experiment is the precise control of a floating linear metal particle introduced into a parallel-plate electrode system. While several studies have developed effective methods for controlling floating metal particles [12–17], these efforts have focused mainly on particles ranging in size from millimeters to centimeters. However, metallic contaminants in practical electrical equipment are typically with diameters in the sub-millimeter scale [18–20]. For better reflecting these conditions, this paper employs a metal particle with a length of 5 mm and a diameter of 0.8 mm. The use of sub-millimeter metal particles imposes more stringent requirements on the control apparatus, including:

(1) The metal particle should maintain an axisymmetric geometry to ensure consistency with the assumptions made in 2D numerical simulations.

(2) To preserve particle's original surface chemical properties and avoid interference with charge relaxation dynamics, no adhesives are applied to the metal particle.

(3) The position of the floating particle should be adjustable with high precision.

(4) Once positioned, the metal particle should remain stable, exhibiting no displacement or rotation under the influence of electrostatic forces.

To meet these requirements, a floating control apparatus is developed, as shown in figure 1. For better reproducibility of experimental results, key mechanical details of the apparatus are provided as below.

(1) A 0.5 mm thick elastic silicone pad with a central through-hole (slightly smaller than the metal particle diameter) is used to achieve an interference fit that clamps the particle via elastic deformation. The pad is tensioned and connected to two identical lifting stages, allowing the particle to remain stably floating in the gap while maintaining axisymmetric geometry.

(2) Two lifting stages, adapted from the micrometer screws, connect to two ends of the silicone pad, and offer positional control with an accuracy of up to 0.01 mm.

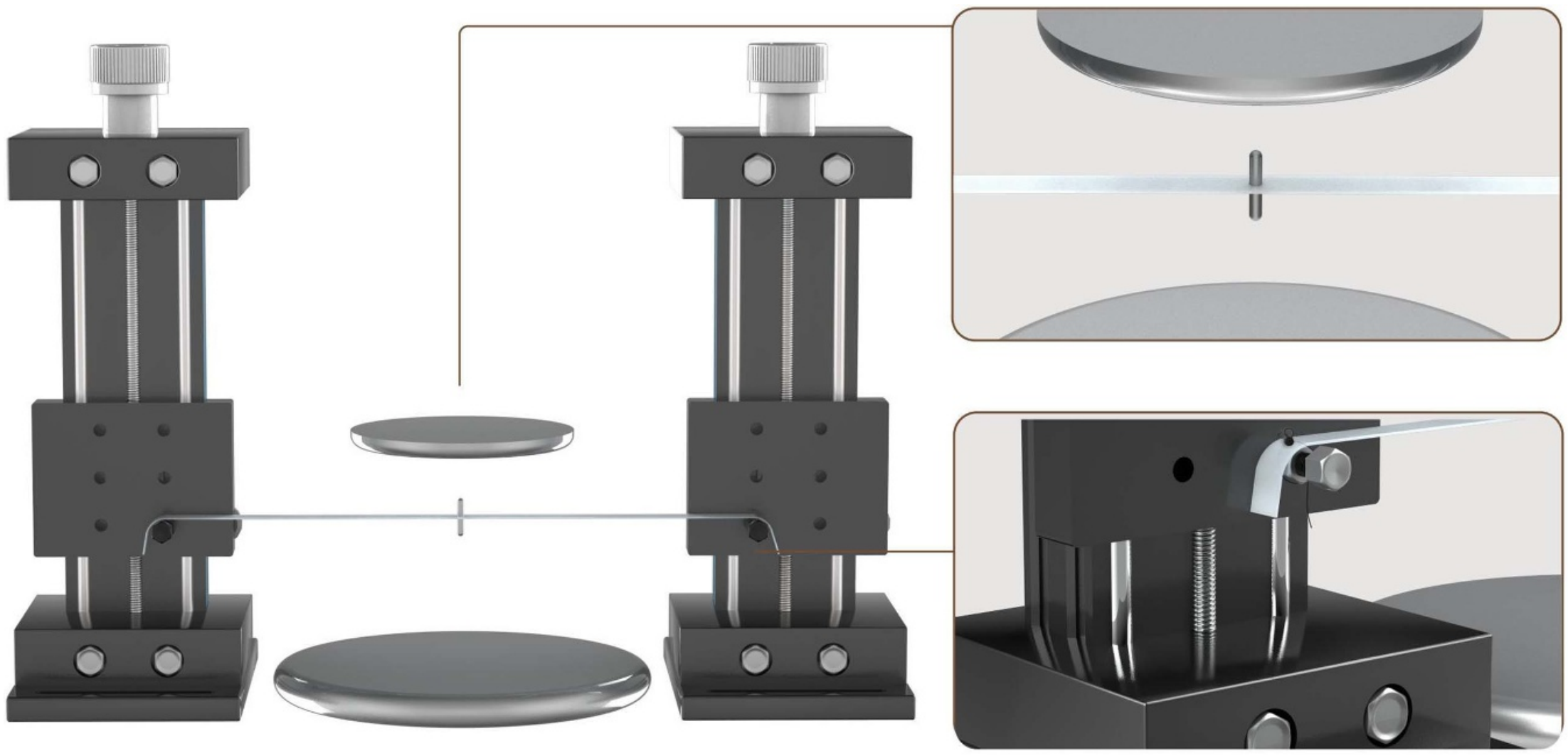

**Figure 1.** Floating control apparatus and key mechanical details involved.

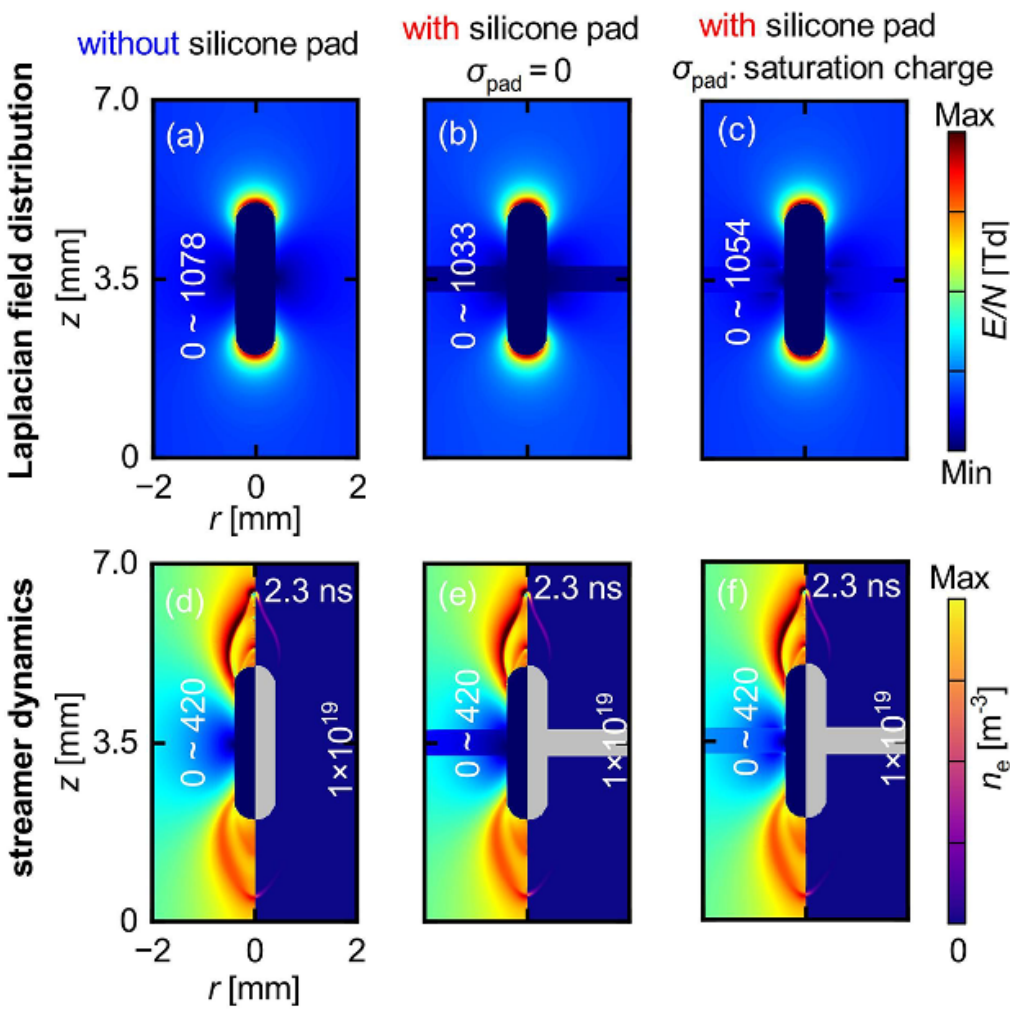

**Figure 2.** Comparison of the impact of the silicone pad on the Laplacian field (a)–(c) and on streamer dynamics (d)–(f).

(3) Polyester threads link the ends of the silicone pad to the positioning screws of the lifting stages. For mechanical stability, the pad is 10 mm wide, with its width plane aligned to the locknut plane and held under tension. This configuration restricts the degrees of freedom of the pad and the metal particle, effectively suppressing displacement and rotation caused by electrostatic forces.

Notably, no partial discharges are observed from the silicone pad in the experiment. Furthermore, simulations are conducted using a reduced-scale geometry to investigate the influence of the silicone pad on both the Laplacian electric field (see figures 2(a)–(c)), and the streamer dynamics (see figures 2(d)–(f)). For the cases with the silicone pad, two initial surface charge density conditions are considered: $\sigma_{\mathrm{pad}} = 0$ and saturation charge. The value of saturation charge is estimated by using the approach proposed by Pedersen and Blaszczyk [21]. The core idea of [21] is that once the normal electric field at the dielectric surface is zero, space charge can no longer migrate to the dielectric, so its surface charge reaches saturation charge condition. By implementing the saturation charge formulation described in reference [21] as a boundary condition on the silicone pad near the particle, the resulting $\sigma_{\mathrm{pad}}$ distribution will self-consistently satisfy the saturation charge condition. The simulation results indicate that the influence of the silicone pad on the discharge is negligible.

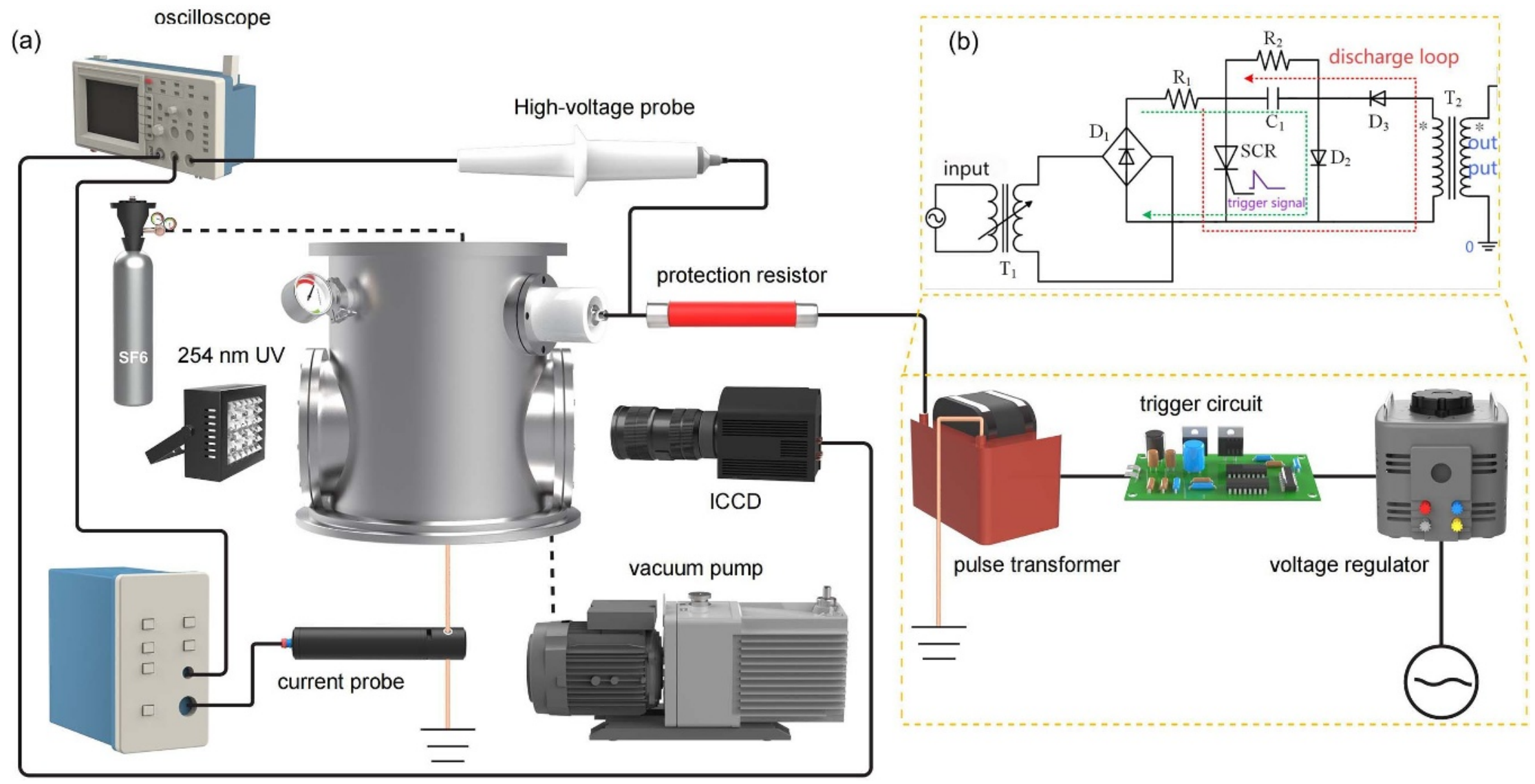

**Figure 3.** (a) Schematic diagram of the experimental system. (b) Circuit diagram of the microsecond pulsed high-voltage source.

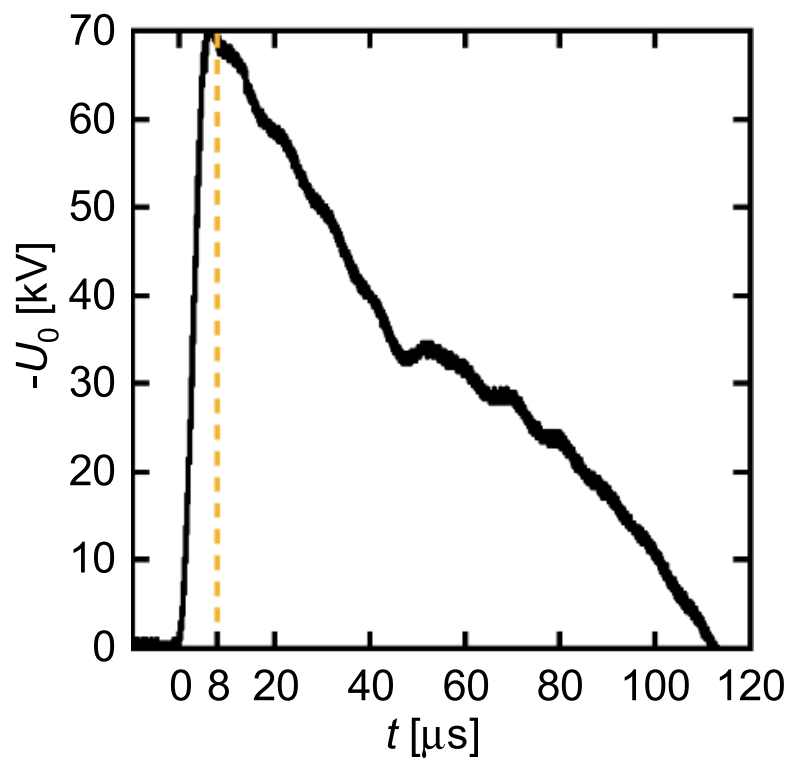

**Figure 4.** Microsecond pulsed high-voltage waveform. Notably, the waveform is plotted with inverted polarity for improved readability.

### 2.2. Experimental system

The complete experimental system is shown in figure 3(a), where the solid black lines represent electrical connections and the dashed black lines represent gas flow paths.

#### *2.2.1. Negative polarity microsecond high-voltage pulse source*

The circuit diagram of the microsecond high-voltage pulse source used in this paper is shown in figure 3(b). A voltage regulator (T1) first charges a storage capacitor, and a trigger circuit then initiates a low-voltage pulse, which is stepped up by a pulse transformer (T2) to generate a high-voltage output.

The low-voltage pulse generation circuit is detailed as below. The voltage regulator (T1) and rectifier bridge (D1) supply a controllable DC voltage to charge capacitor C1, with resistor R1 limiting the current. Diodes D2 and D3 isolate the charging and discharging paths, preventing premature triggering of the transformer. Resistor R2 discharges C1 between cycles, enabling a fast voltage response. Its resistance is set significantly higher than that of R1 to maintain sufficient voltage on C1. A microcontroller unit triggers the silicon-controlled rectifier to initiate the pulse. The resulting microsecond high-voltage waveform applied to the high-voltage electrode is shown in figure 4, featuring a peak voltage of $-70$ kV and a rise time of 8 $\mu$s. All experiments in this paper employ single-pulse discharges.

#### *2.2.2. Optical diagnostics*

An ICCD camera (TRC411-H20-U, Intelligent Scientific Systems Co.) equipped with a long-focus lens (105 mm f2.8, Sigma EX) is used to capture optical photographs of the discharge process. Photos of the streamer discharges are obtained in gate mode with an exposure time of 0.5 ns. The optical gain is fixed in this paper. The ICCD camera is triggered by a signal from the microcontroller unit. The ICCD images undergo background noise reduction and contrast enhancement to highlight the discharge regions.

*2.2.3. Vacuum chamber*

All experiments are performed at room temperature inside a vacuum chamber. The chamber is first evacuated to 1 Pa using a vacuum pump, then filled with 99.99% $SF_6$ gas to 1 atm, with a pressure uncertainty of approximately 0.1%.

*2.2.4. Gap electrode*

The electrode contains a pair of parallel-plates with the particle introduced into the gap as shown in figure 1. The upper plate is the high-voltage electrode, and the lower plate is the grounded electrode. The combined gap distance $d$ between the high-voltage and grounded electrodes is 17 mm. To mitigate edge-induced electric field distortions, both plate edges are chamfered. Additionally, the high-voltage electrode is designed with a smaller radius than the grounded electrode to prevent unintended breakdown propagating through the lifting stages.

The floating metal particle has a length of 5 mm and a diameter of 0.8 mm. Both tips of the metal particle are polished into smooth hemispheres with a diameter equal to that of the particle (0.8 mm). The particles are fabricated via fixture-attached CNC machining, and tip geometry is verified by optical microscopy. In our experiments, each metal particle is limited to at most 20 breakdown discharges and is then replaced with a new one. We consider that this approach can avoid surface oxidation of the metal particle, and thereby ensure that the charge transfer mechanism of metal particle remains dominated by electrostatic induction rather than polarization. The short gap $g_s$ between particle's top tip and the high-voltage electrode is 4 mm, while the long gap $g_l$ between particle's bottom tip and the grounded electrode is 8 mm. The discharge electrode is placed inside the vacuum chamber.

*2.2.5. Electrical diagnostics*

The discharge current is measured using a active current probe (TCPA300, Tektronix), and the high-voltage waveform is measured using a high-voltage probe (P6015A, Tektronix).

*2.2.6. Oscilloscope*

All signals (voltage, current and ICCD gate) are recorded on the same oscilloscope (MDO3034, Tektronix), ensuring a common time base.

### 2.3. Discharge triggering scheme

In this paper, continuous 254 nm ultraviolet (UV) illumination is applied to the discharge gap to simulate pre-ionization at both ends of a floating metal particle prior to breakdown [14, 22, 23]. A single-triggered pulse discharge is employed, with a minimum interval of 10 s between two trigger signal to ensure a relatively stable charging state of the metal particle.

Despite this triggering scheme, the initial charge on the particle remains difficult to control precisely, resulting in considerable uncertainty in the discharge onset-approximately 800 ns. This uncertainty poses a major challenge, as it exceeds the typical duration of the streamer development phase. To reliably capture streamer morphologies under such uncertainty, extensive repeated trials are required, making the experimental process time-consuming.

## 3. Error analysis and rationality assessment

Due to the inability to precisely control the particle's initial charge and the technical limitations of the pulsed voltage source, the experiment exhibits limited discharge repeatability. This constrains the ability to perform accurate time-resolved sequences of the discharge process. However, the experimental setup still can meet the research objectives, namely, capturing the representative streamer morphologies. A detailed error analysis and rationality analysis are discussed as below.

### 3.1. Error analysis

*3.1.1. Timing jitter*

Although 254 nm UV pre-ionization reduces initial charge variability, the initial charge cannot be precisely controlled. Variations in initial charge will inevitably alter the Laplacian field around the particle, introducing $\sim$800 ns timing jitter.

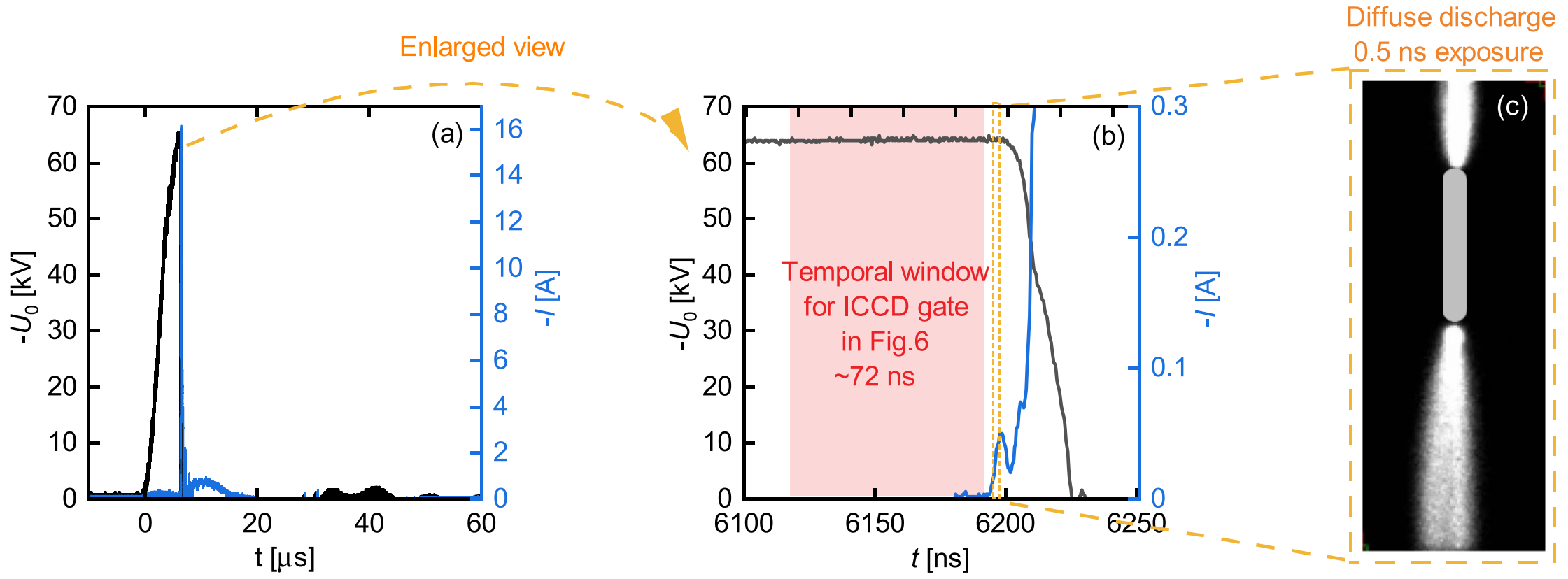

**Figure 5.** (a) Measured voltage and current waveforms. (b) Enlarged view of voltage and current waveforms at breakdown moment, along with the ICCD gate temporal window. (c) Diffuse discharge photograph captured at breakdown moment. Notably, all waveforms are plotted with inverted polarity for improved readability. Panels (a) and (b) are obtained from two separate discharge events with similar breakdown voltage.

#### *3.1.2. Breakdown voltage jitter*

Due to the inherently high electric field required for $SF_6$ breakdown, $SF_6$ streamers tend to initiate and propagate rapidly. To extend the streamer propagation time, a relatively large electrode gap ($d = 17$ mm) is used. This increases the required breakdown voltage [24], which reaches -64 kV under microsecond-pulse conditions, as shown in figure 5(a). However, a nanosecond pulse generator capable of delivering the required voltage amplitude is not available due to the limited conditions of the authors. Therefore, a microsecond pulse source ($-70$ kV peak, 8 $\mu$s rise time, see figure 4) is used as an alternative. In this configuration, breakdown occurs along the rising edge of the voltage pulse (see figure 5(a)). Given a timing jitter of $\sim$ 800 ns, this translates to a theoretical voltage variation of around 6 kV. In practice, the actual breakdown voltage jitter for the captured streamer morphologies of figure 6 is $\sim$ 1.2 kV.

In summary, we must acknowledge that our experimental setup cannot provide accurate time-resolved characterization.

### 3.2. Rationality assessment of streamer morphology

The rationality lies in the facts that the reported photographs in figure 6 reflect the realistic streamer morphologies. This is justified by two points:

#### *3.2.1. Temporal window*

It should be acknowledged that in some nanosecond-pulse breakdown experiments [25–28], the current may match streamer propagation. But in our microsecond-pulse breakdown experiments, the current during the streamer stage shows irregular oscillations and lacks characteristic features for identifying streamer propagation. This is possibly due to RLC oscillations in the discharge circuit, similar to the analysis in [29]. Nonetheless, diffuse discharge consistently occurs at the moment of breakdown, marked by a sharp voltage drop and current peak (see figure 5(b)). All photographs reported in figure 6 are taken within a temporal window of $\sim$ 72 ns before the breakdown moment. This temporal window falls well within the typical streamer timescale, confirming that the recorded photographs reflect streamer morphologies.

#### *3.2.2. Optical intensity of the photographs*

After breakdown, the discharge will first transition into a diffuse discharge mode, characterized by a significant high optical intensity (see figure 5(c)). In contrast, the optical intensity of the photographs in figure 6 is rather lower, indicating that they correspond to the streamer stage.

## 4. Results and Discussion

### 4.1. Weak optical intensity of $SF_6$ streamer

Despite the ICCD camera being set to a sufficiently high optical gain, the optical intensity of the $SF_6$ streamer in figure 6 remains rather weaker than that of air streamers recorded under identical 0.5-ns exposure conditions in [30]. According to [31], this weak intensity is attributed to the high molecular weight and specific structure of $SF_6$, which lead to efficient non-radiative quenching of excited states. Further, comparative analysis between $SF_6$ and $N_2$ in [31] confirms the rationality of lower optical intensity of $SF_6$ streamers. Since the objective of this paper is to compare with the findings of [5], it is necessary to use pure $SF_6$ as the discharge gas to maintain consistency. To improve spatial resolution, future work may adopt several refined strategies. For example, $N_2$ may be introduced as a tracer gas [25, 32, 33]. In addition, high-resolution electric field measurement might be employed to characterize streamer dynamics [34, 35].

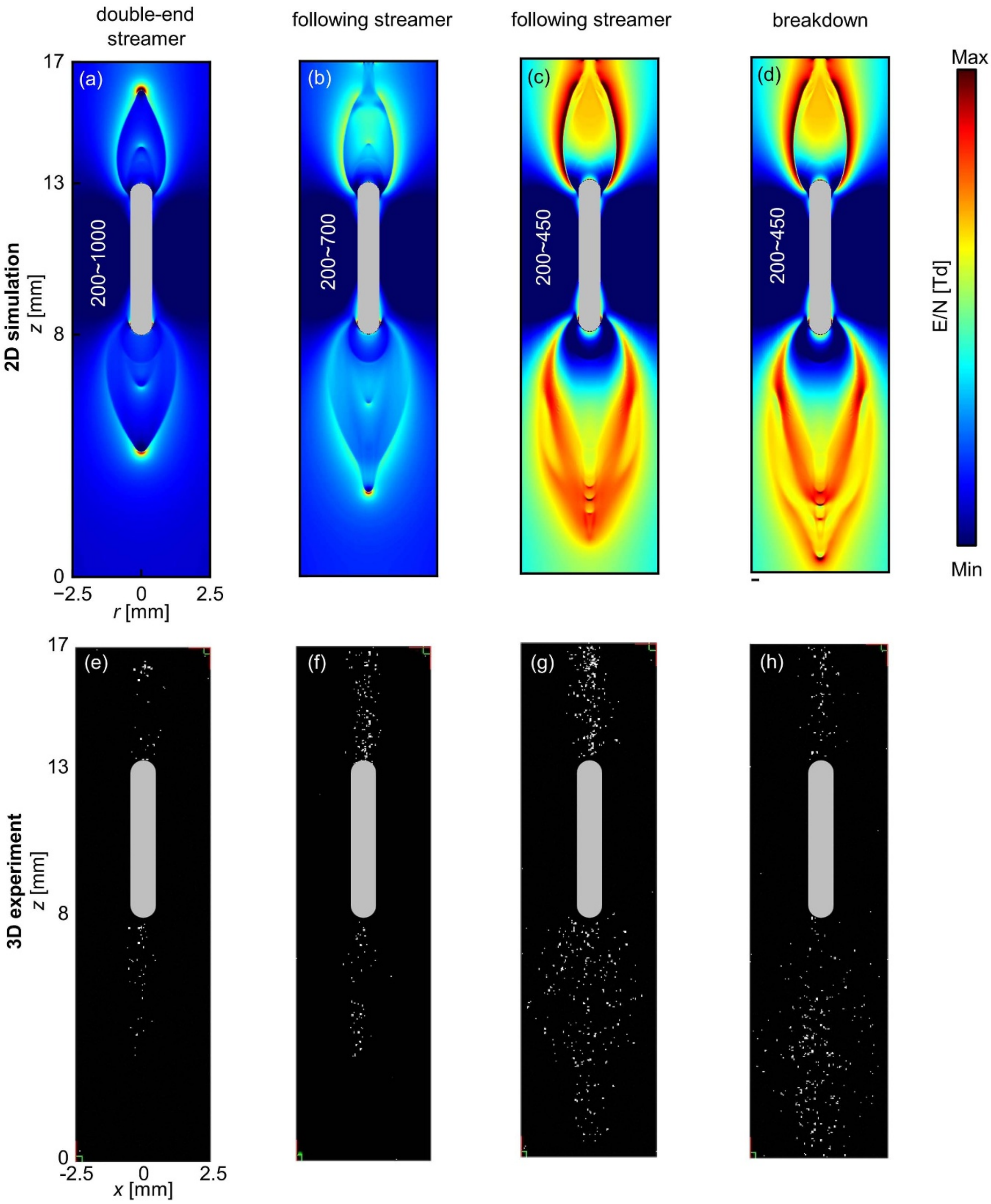

**Figure 6.** (a)–(d) Reduced electric field profile of 2D axisymmetric simulation. (e)–(h) Optical photographs of 3D experiment.

### 4.2. Double-end streamer inception

***Consistencies*** Both experimental and simulation results support the double-end streamer inception proposed in [5]. Specifically in experiments (see figure 6(e)), the discharge photograph reveals luminous regions both in the upper and lower gaps, indicating streamer initiation at both tips of the floating metal particle. Simulation results (see figure 6(a)) demonstrate the presence of forward ionization fronts in both gaps, i.e. double-end streamers. In particular, the streamer in the upper gap has not bridged.

These features confirm that discharge in the longer gap does not rely on the short-gap bridging; namely, there is no deterministic relationship between the two processes.

The double-end streamer inception is governed by a synergistic mechanism involving electron charging during the avalanche phase in the short gap and the electrostatic induction of the metal particle-both of which are indispensable. Specifically, in the negative applied field, the above-mentioned electrons drift rapidly toward the top tip of the metal particle. Combing with electrostatic induction, the metal particle redistributes surface charge on a timescale of $\sim 10^{-19}$ s. This results in the transfer of negative charge from the particle's top tip to the bottom tip, enhancing the electric field near the bottom tip. The enhanced electric field continues to intensify and, within an extremely short timescale-prior to the bridging of the short gap-reaches the streamer threshold, ultimately triggering the streamer inception in the long gap.

This phenomenon is a characteristic feature of metal-particle-induced-discharge and remains rarely observed in the context of floating dielectric particles (see Mirpour and Nijdam [12]). A possible explanation lies in the fundamental difference in their charge transfer mechanisms. In dielectric particles, charge transfer primarily occurs through polarization, which is a localized mechanism that cannot achieve global redistribution across the whole particle surface. This limitation reduces their ability to enhance the local electric field in the longer gap within a short timescales. In contrast, in metal particles, charge transfer occurs through electrostatic induction of free electrons, which enables rapid, global redistribution of surface charge, allowing real-time modulation of the local electric field in the long gap.

### 4.3. Formation and propagation of following streamers

Note that, the sub-figure in figure 6 represents only still photograph per discharge, without time-resolved evolution between them as acknowledged in section. 3. So what we can robustly see is a 3D branching streamer tree at different stages, but they fall short of constituting a strict validation for the *Following Streamer* mechanism.

Here, within our current conditions, we provide indirect inference of the *Following Streamer* mechanism based on the luminous phenomena within streamer channel. Specifically, for strong electronegative insulating gases such as $SF_6$ and $CO_2$, the electron density within the streamer channel is significantly reduced due to attachment. As a result, excitation reactions within the channel cannot be effectively sustained, leading to negligible luminous intensity within streamer channel, as shown in figure 6(e) of this paper. Similar phenomena was also observed in $CO_2$ streamers, for example, in figure 8(f) of Omori *et al* [25] and in figure 8(a) of Li *et al* [36], no obvious luminescence is observed within the streamer channel. However, in figures 6(f)–(h) of this paper, a remarkably different luminous phenomenon within the streamer channel is observed in the long gap. Namely that, the streamer channel in the long gap exhibits strong luminous phenomenon, suggesting the occurrence of new ionization inside the channel, which may indirectly infer the formation of subsequent following streamers. Therefore, both experimental and simulation results in figure 6 are qualitatively consistent with the general idea of *Following Streamer* proposed in [5]. However, notable discrepancies arise in the number and propagation path of the following streamers. For simplicity, in the following discussions, the term "streamer" will exclusively refer to those occurring in the long gap, excluding those in the short gap.

#### *4.3.1. Consistencies*

Specifically in experiments (see figures 6(f) and (g)), the primary streamer has propagated away from the bottom tip of the metal particle. Meanwhile, newly emerging localized luminous regions appear near the same tip, indicating the formation of new streamers. In simulations (see figures 6(b) and (c)), new ionization waves are observed to originate from the bottom tip of the particle and propagate forward, signifying the formation of new streamers. These newly formed discharges are indirectly inferred as following streamers, qualitatively consistent with the general idea of *Following Streamer* mechanism.

The formation of following streamers is governed by two key aspects: (1) the interaction between the space charge within the plasma and the metal particle, and (2) the electrostatic induction effect of the entire metal particle-both of which are indispensable. Specifically, under negative applied voltage, the negative space charge in the short gap (dominated by electrons, which contribute six times more than negative ions) is transported toward the top tip of the particle, while the positive ions in the long gap, is transported toward the bottom tip of the particle. To maintain electrostatic equilibrium, the particle's surface charges undergo real-time redistribution governed by electrostatic induction, leading to an increase of negative surface charge at the bottom tip. This charge redistribution enhances the local electric field near the bottom tip, making the effect of the metal particle similar to the rising edge of a pulsed voltage. Consequently, the conditions become favorable for the formation of following streamers in the long gap.

*4.3.2. Discrepancies*

(1) Number of the following streamers

Specifically in experiments (see figure 6(g)), two distinct types of discharge morphology are observed in the long gap. The first type corresponds to the primary streamer, which positions away from the bottom tip of the metal particle, and maintains a narrow luminous channel characteristic, which is considered a single-streamer structure. The second type corresponds to the following streamers, positions near the bottom tip of the particle. In these cases, the luminous region appears broader and exhibits an optical intensity comparable to that of the primary streamer, indicating the formation of multiple new streamers-i.e. multiple following streamers. These experimental observations differ from the simulation results (see figure 6(c)), in which only a single following streamer is generated each time, and it consistently forms on the axis of symmetry.

This discrepancy arises from the spatial locations where the following streamers form. In 3D scenarios, new streamers may originate from positions offset from the central axis [37–40], thereby breaking the symmetry of the electric field. As a result, the number of following streamers cannot be fully described by the 2D axisymmetric coordinate system.

(2) Propagation path of the following streamers

Since the formation of following streamers inherently breaks the symmetry of the electric field, this asymmetry is further amplified during their subsequent propagation [41–43]. Specifically in experiments (see figure 6(g)), following streamers may deviate from the central axis and adopt new propagation paths that curve around the primary streamer, resulting in a significantly broader luminous region compared to that of the primary streamer. Furthermore, interactions among newly formed following streamers and with pre-existing ones may break their individual stability [44–46], further contributing to deviations from axisymmetric propagation. These experimental observations contrast with simulation results (see figure 6(c)), in which following streamers consistently propagate along the axis of symmetry. Consequently, the propagation behavior of following streamers cannot be fully captured within a 2D axisymmetric coordinate system.

### 4.4. Streamer bridging of the combined gap

As shown in figure 6(h), once the luminous intensity extends across the entire combined gap, the discharge channel becomes diffuse, and the distinction between the primary and following streamers becomes indistinct. This observation suggests that multiple streamers have reached the grounded electrode, indicating that streamer bridging has been completed across the entire combined gap.

## 5. Conceptual model for following streamer mechanism

To investigate $SF_6$ streamer propagation triggered by floating linear metal particles under a negative applied voltage, a conceptual model is proposed as shown in figure 7, featuring both a 2D axisymmetric schematic and a 3D Cartesian schematic, as well as an explanation of the discrepancies between these two schematics.

In the 2D axisymmetric schematic, primary streamers are initiated simultaneously from both ends of the metal particle-demonstrating a double-end inception phenomenon. Subsequently, the following streamers form and propagate along the symmetry axis in the long gap, ultimately resulting in the bridging across the entire combined gap.

In the 3D Cartesian schematic, double-end streamer inception also occurs, but the behavior of following streamers differs markedly. Rather than being confined to the symmetry axis, new streamers may form at off-axis positions, allowing multiple following streamers to develop simultaneously. Their propagation paths often deviate from the symmetry axis, ultimately resulting in bridging of the entire combined gap.

These discrepancies between the 2D and 3D schematic can be attributed to three key factors: (1) the off-axis formation of following streamers; (2) the breaking of electric field symmetry; and (3) interactions between multiple streamers during propagation. Together, these elements highlight the limitations of 2D axisymmetric models in accurately capturing the complex spatial dynamics of streamer development in the context of metal-particle-induced-discharge.

In our simulations, only varying the particle length (to 2 mm) or the particle diameter (to 0.2 mm) does not fundamentally affect the occurrence of double-ended inception or the following streamer.

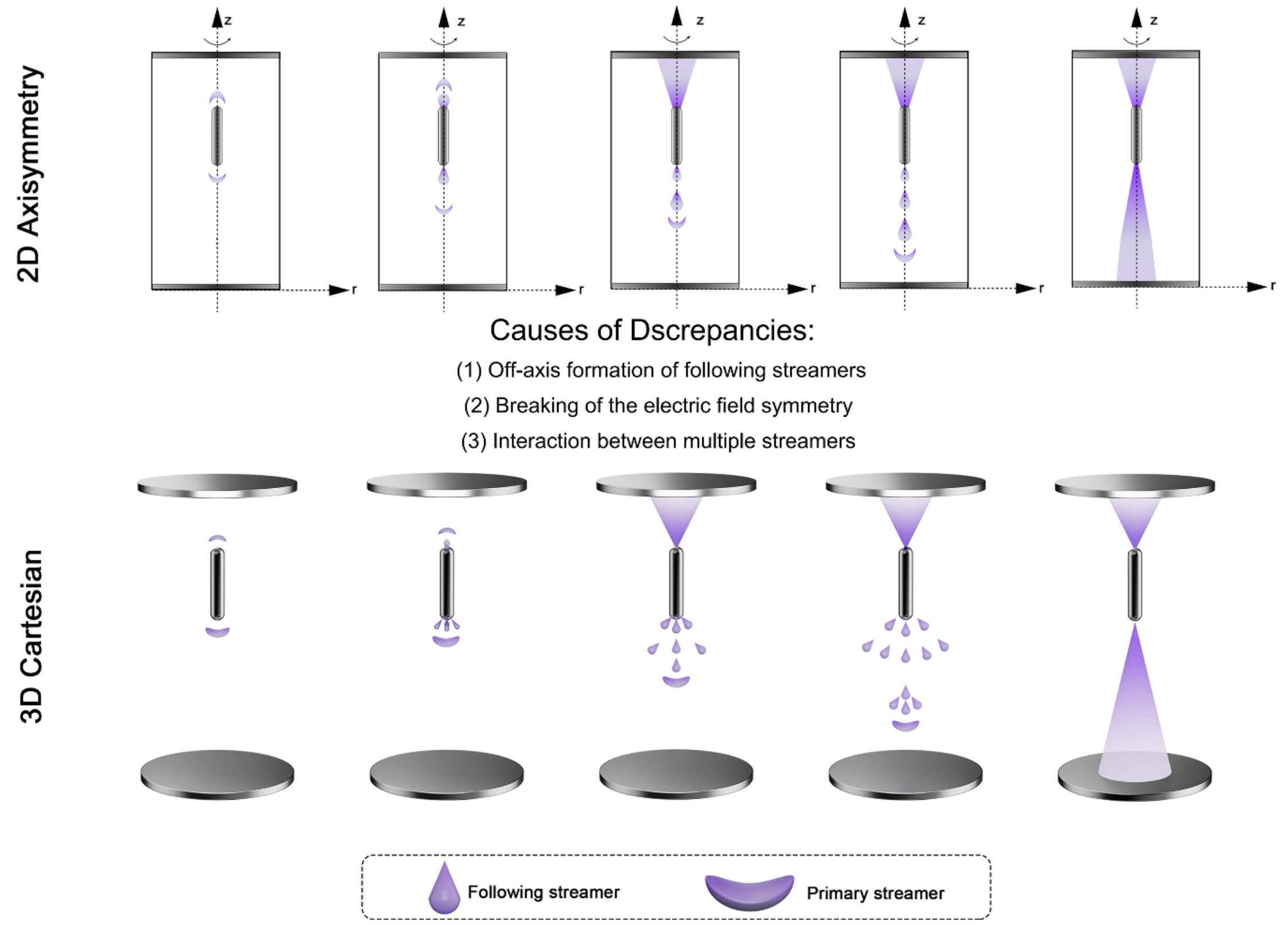

**Figure 7.** Conceptual model for *Following Streamer* mechanism under negative applied voltage.

Therefore, within the above-mentioned parameter range, the conceptual model is considered applicable. However, its validity for smaller particles and under positive applied voltage conditions needs to be examined in future studies.

## 6. Conclusions and outlook

### 6.1. Conclusions

To further study the *Following Streamer* mechanism proposed in [5], we conducted both experimental and numerical investigations of $SF_6$ streamer propagation induced by a floating linear metal particle under a negatively applied voltage. The results from 3D experimental observations are compared with 2D axisymmetric simulations to assess the consistency and discrepancies in describing the fundamental features of the *Following Streamer* mechanism. The key findings are as follows:

***Consistencies.*** The general idea of the *Following Streamer* mechanism is indirectly inferred: (1) Double-end streamer inception (2) Formation of following streamers

***Discrepancies.*** In the 3D scenarios, following streamers form at off-axis locations, breaking the symmetry of the electric field. Additionally, interactions among multiple streamers further enhance this asymmetry. As a result, 2D axisymmetric fluid simulations are unable to fully capture: (1) The number of following streamers (2) Their off-axis propagation paths

Finally, we propose a conceptual model that incorporates both a 2D axisymmetric schematic and a 3D Cartesian schematic. This model provides a more comprehensive framework for understanding the $SF_6$ streamer propagation in the presence of floating metal particles.

### 6.2. Outlook

This paper falls short of constituting a strict quantitative validation for the *Following Streamer* mechanism, so it still requires further investigation. For example, future work may be valuable to employ advanced experimental technology [47, 48] to capture time-resolved streamer evolution, quantitative comparison of charge densities, electric fields, or propagation velocities, nor statistical analysis of occurrence frequency.

## Acknowledgments

The authors gratefully acknowledge funding support from the National Natural Science Foundation of China (Grant No. 52407176), the Postdoctoral Fellowship Program of CPSF (Grant No. GZB20230326), and State Grid Beijing Electric Power Company (52020226000G).

## Data availability statement

All data that support the findings of this study are included within the article (and any supplementary files).

## Conflict of interest

The authors have no conflicts to disclose.

## Appendix. Numerical scheme

***2D Axisymmetric geometry and boundaries.*** The general geometry and boundary conditions are consistent with those in [5], except that the specific dimensions have been adjusted to match the experimental setup in this paper, as described in section 2.2. Notably, the boundary conditions applied to the floating metal particle should not only accurately reflect the local field enhancement, but also accurately reflect the electrostatic induction of the metal, namely, a self-consistent redistribution of surface charges across the entire particle during every computational time step.

***Governing equations.*** The fluid model under the local mean energy approximation is employed, consistent with those in [5]. Besides, the following improvements have been made: the chemical species are adjusted to include $e$, $SF_5^+$, $SF_4^+$, $SF_3^+$, $SF_2^+$, $SF^+$, $S^+$, $F^+$, $SF_6^-$, $SF_5^-$, $F^-$, $SF_6$, $SF_5$, F, and $SF_6^*$. Here, $SF_6^*$ represents multiple excitation states, treated as a single species. The corresponding cross-sections are all taken from the Biagi database [49]. The photoionization source terms in the continuity equations are calculated using the computational photoionization model and three-group Helmholtz parameters for 1 atm $SF_6$ proposed in [50], in which the theoretical foundation is based on [51, 52], while the numerical computation formulates using a set of Helmholtz equations similar to [53, 54]. The reaction source terms are extended to incorporate the contributions of detachment. The value of detachment coefficient is taken from figure 41 of [31].

***Grid spacing.*** A carefully designed unstructured grid scheme is employed, and the minimum grid spacing is 0.1 $\mu$m. The simulations are performed using 2×Intel® Xeon® Gold 6246 R @ 3.4 GHz and 12 memory channels, each equipped with 64 GB of DDR4-2933 RAM.

***Time step.*** The implicit backward differentiation formula (BDF) method is used, with a maximum BDF order of 2 and a minimum BDF order of 1. The relative tolerance is set to $10^{-2}$. A parallel sparse direct solver is selected as the direct linear system solver. Notably, the maximum time step is recommended 0.5 ps, and the rationale will be presented elsewhere.

## ORCID iDs

Zihao Feng 0009-0008-0015-6837
Liyang Zhang 0000-0001-9909-9263